\documentclass[a4paper,11pt]{article}
\usepackage{pos}

\usepackage{amsfonts} 
\usepackage{amsmath} 
\usepackage{graphicx} 
\usepackage{array} 
\usepackage{bm} 
\usepackage{latexsym} 
\usepackage{longtable} 
\usepackage{hyperref} 
\usepackage{multirow}
\usepackage{xcolor}
\usepackage{bbold}
\usepackage{subcaption}
\usepackage{lipsum}
\usepackage{mathtools}
\usepackage{comment}
\usepackage{paralist}
\usepackage{placeins}
\usepackage{verbatim}

\graphicspath{{figs/}}

\providecommand{\ket}[1]{\lvert#1\rangle}
\providecommand{\mate}[3]{\langle#1\lvert#2\rvert#3\rangle}

\providecommand{\innerp}[2]{\langle#1\vert#2\rangle}

\definecolor{HLBlue}{HTML}{6599FF}
\definecolor{HLOrange}{HTML}{FF6600}

\newcommand{\GxOne}{[\gamma_5\otimes\mathbb{1}]}

\newcommand{\red}[1]{\textcolor{red}{#1}} 
\newcommand{\blue}[1]{\textcolor{blue}{#1}}

\title{Deep learning study on the Dirac eigenvalue spectrum of staggered quarks}

\author[a,1]{Hwancheol Jeong}
\author[b,1]{Chulwoo Jung}
\author[c,1]{Seungyeob Jwa}
\author[c,1]{Jeehun Kim}
\author[d]{Nam Soo Kim}
\author[c,1]{Sunghee Kim}
\author*[c,1]{Sunkyu Lee}
\author[c,1]{Weonjong Lee}
\author[e]{Youngjo Lee}
\author[c,1]{Jeonghwan Pak}
\author[c,1]{Chanju Park}

\affiliation[a]{
  Department of Physics,
 	Indiana University, Bloomington, IN 47405, USA 
}

\affiliation[b]{
  Physics Department, 
  Brookhaven National Laboratory, Upton, New York 11937, USA
}

\affiliation[c]{
  Lattice Gauge Theory Research Center, FPRD, and CTP, 
  Department of Physics and Astronomy, \\
  Seoul National University, Seoul 08826, South Korea
}

\affiliation[d]{
	Institute of New Media and Communications, 
	Department of Electrical and Computer Engineering, \\
	Seoul National University, Seoul 08826, South Korea
}

\affiliation[e]{
	Department of Statistics,
	Seoul National University, Seoul 08826, South Korea						 
}

\note{For the SWME collaboration}

\emailAdd{sunkyu131@snu.ac.kr}
\emailAdd{wlee@snu.ac.kr}

\abstract{ 
We study the chirality of staggered quarks on the Dirac eigenvalue
spectrum using deep learning (DL) techniques. 
The Kluberg-Stern method to construct staggered bilinear operators
conserves continuum property such as recursion relations, uniqueness
of chirality, and Ward identities, which leads to a unique and
characteristic pattern (we call it ``leakage pattern (LP)'') in the
matrix elements of the chirality operator sandwiched between two quark
eigenstates of staggered Dirac operator.
DL analysis gives $99.4(2)\%$ accuracy on normal gauge configurations
and $0.998$ AUC (Area Under ROC Curve) for classifying non-zero mode
octets in the Dirac eigenvalue spectrum.
It confirms that the leakage pattern is universal on normal gauge
configurations. 
The multi-layer perceptron (MLP) method turns out to be the best DL
model for our study on the LP.
}

\FullConference{ The 38th International Symposium on Lattice Field
  Theory, LATTICE2021 26th-30th July, 2021 Zoom/Gather@Massachusetts
  Institute of Technology }

\begin{document}
\maketitle

\section{Introduction}
Here, we present technical details on how to do a deep learning
analysis on low non-zero modes for staggered Dirac operators.
This work is a follow-up paper on our previous papers: Ref.~\cite{
  Cundy:2016tmw, Jeong:2017kst, Jeong:2020map, SWME:2020yip}.
\section{Staggered Dirac eigenstate}
Here, we adopt the same notation as in Ref.~\cite{ SWME:2020yip}. 
Staggered Dirac operator, $D_s$, is anti-Hermitian ($D_s^\dagger =
-D_s$) and its eigenvalues are pure imaginary.
\begin{equation}
  D_s\ket{f_i} = i\lambda_i\ket{f_i}
  \label{eq:eigen_eq_serial}
\end{equation}
where $\lambda_i$ is a real eigenvalue.
Here, the eigenstates are normalized such that $\innerp{f_i}{f_k} =
\delta_{ik}$.
Under conserved $U(1)_A$ axial transformation $\Gamma_\epsilon$, Dirac
operator transforms into its Hermitian conjugate,
\begin{equation}
  \Gamma_\epsilon D_s \Gamma_\epsilon = D^\dagger_s = -D_s.
\end{equation}
And an eigenstate $\ket{f_i}$ is related with its parity partner
$\ket{f_{-i}}$ as follows,
\begin{align}
  D_s\Gamma_\epsilon\ket{f_i} &= -i\lambda_i\Gamma_\epsilon\ket{f_i}, \\
  D_s\ket{f_{-i}} &= -i\lambda_i\ket{f_{-i}}.
\end{align}
Hence,
\begin{alignat}{3}
  \Gamma_\epsilon\ket{f_i} &= e^{i\theta}\ket{f_{-i}}
  \quad &\longleftrightarrow \quad &
  \Gamma_\epsilon\ket{f_{-i}} = e^{-i\theta}\ket{f_{+i}}.
\end{alignat}
The eigenvectors can be classified into two categories: one is a set
of zero modes and the other is a set of non-zero modes.
At the finite lattice spacing, there is no exact zero mode
($\lambda_i=0$) with staggered fermions \cite{ Smit:1986fn}.
Those eigenmodes which correspond to zero modes in the continuum, are
named ``would-be zero modes''.
For more details, refer to Ref.~\cite{ SWME:2020yip}.
%
%

%
\section{Kluberg-Stern method}
In general, there are two independent methods to transcribe a
continuum operator to its lattice version using staggered fermions:
one is the Golterman (Gol) method \cite{ Golterman:1984cy} and the
other is the Kluberg-Stern (Klu) method \cite{ Kluberg-Stern:1983lmr}.
The Klu method has a number of advantages in the study on staggered
Dirac eigenvalue spectrum, compared with the Gol method, since it
satisfies (a) recursion relations, (b) uniqueness of chirality, and
(c) Ward identities.
In the Klu method, a quark bilinear operator with spin $S$ and taste
$T$ is
\begin{equation}
  \mathcal{O}_{S\times T}(x)
  \equiv \bar{\chi}(x_A)[\gamma_S\otimes \xi_T]_{AB} \chi(x_B) 
  = \bar{\chi}(x_A)\overline{(\gamma_S\otimes \xi_T)}_{AB} U_{AB}\chi(x_B)
\end{equation}
where $\chi$ is a staggered quark field, $x_A=2x+A$, $x$ is the
coordinate of a hypercube, and $A_\mu,B_\mu\in\{0,1\}$.
Here, $\overline{(\gamma_S\otimes \xi_T)}_{AB} =
	\dfrac{1}{4}\text{Tr}\left(\gamma^\dagger_A\gamma_S\gamma_B\gamma^\dagger_T\right)$.
The $U_{AB}$ is inserted to make the operator gauge-invariant,
\begin{equation}
  U_{AB} \equiv
  \mathbb{P}_{SU(3)}
  \left[ \sum_{p\in \mathcal{C}}V(x_A, x_{p_1}) V(x_{p_1}, x_{p_2})
    \cdots V(x_{p_n},x_B)\right]
\end{equation}
where $\mathbb{P}_{SU(3)}$ is the $SU(3)$ projection, $\mathcal{C}$
denotes the complete set of shortest paths from $x_A$ to $x_B$, and
$V(x,y)$ is HYP-smeared fat link \cite{
Hasenfratz:2001hp}.\footnote{Here, we assume that we use the HYP
staggered action \cite{ Hasenfratz:2001hp} for valence quarks.}
The matrix elements for the chirality operator $\GxOne$ and the shift
operator $[\mathbb{1}\otimes\xi_5]$ are
\begin{align}
  \Gamma_5^{i,k} &= \Gamma_5(\lambda_i,\lambda_k)
  = \mate{f_i}{[\gamma_5\otimes\mathbb{1}]}{f_k} 
  \equiv \sum_{x,A,B}[f_i(x_A)]^\dagger
  \overline{(\gamma_5\otimes\mathbb{1})}_{A,B}U_{AB}f_k(x_B),
  \\
  \Xi_5^{i,k} &= \Xi_5(\lambda_i,\lambda_k)
  = \mate{f_i}{[\mathbb{1}\otimes\xi_5]}{f_k} 
  \equiv \sum_{x,A,B}[f_i(x_A)]^\dagger
  \overline{(\mathbb{1}\otimes\xi_5)}_{A,B}U_{AB}f_k(x_B).
\end{align}
Unlike the Gol operators, the Klu operators satisfy the following:
\begin{itemize}
\item[(a)] recursion relations
  \begin{equation}
    \GxOne^{2n+1} = \GxOne, \quad \GxOne^{2n} = [\mathbb{1}\otimes\mathbb{1}],
  \end{equation}
\item[(b)] uniqueness of chirality, which is natural consequence of
  the above recursion relations,
\item[(c)] Ward identity, 
  \begin{align}
    & \Gamma_\epsilon = [\gamma_5 \otimes \xi_5] 
    = [\gamma_5 \otimes \mathbb{1}] [\mathbb{1} \otimes \xi_5] 
    = [\mathbb{1}\otimes\xi_5][\gamma_5\otimes \mathbb{1}],
    \label{eq:U(1)A_op}
    \\
    & \Gamma_\epsilon [\gamma_5 \otimes \mathbb{1}]
    = [\gamma_5 \otimes \mathbb{1}] \Gamma_\epsilon
    = [\mathbb{1}\otimes\xi_5],
    \label{eq:wi-2}
    \\
    & \Gamma_\epsilon [\mathbb{1}\otimes\xi_5]
    = [\mathbb{1}\otimes\xi_5] \Gamma_\epsilon
    = [\gamma_5 \otimes \mathbb{1}].
    \label{eq:wi-3}
  \end{align}
  Using the Ward identity in Eqs.~\eqref{eq:wi-2} and \eqref{eq:wi-3},
  we can derive the following results:
  \begin{equation}
    \blue{|\Gamma_5^{i,k}|} = |\Xi_5^{-i,k}|  = |\Xi_5^{i,-k}| =
    \blue{|\Gamma_5^{-i,-k}|} = \blue{|\Gamma_5^{k,i}|} 
    = |\Xi_5^{-k,i}| = |\Xi_5^{k,-i}| = \blue{|\Gamma_5^{-k,-i}|}~.
    \label{eq:Ward_id}
  \end{equation}
\end{itemize}
For more details on this, refer to Ref.~\cite{SWME:2020yip}.
Here, we choose $|\Gamma_5^{i,k}|$ to study leakage patterns, because
$|\Xi_5^{i,k}|$ is just a mirror image of $|\Gamma_5^{i,-k}|$.
%

%
%

%
\section{Quartet structure}
%
\begin{figure}[t!]
  \centering
  \vspace*{-5mm}
  \includegraphics[width=0.75\linewidth]{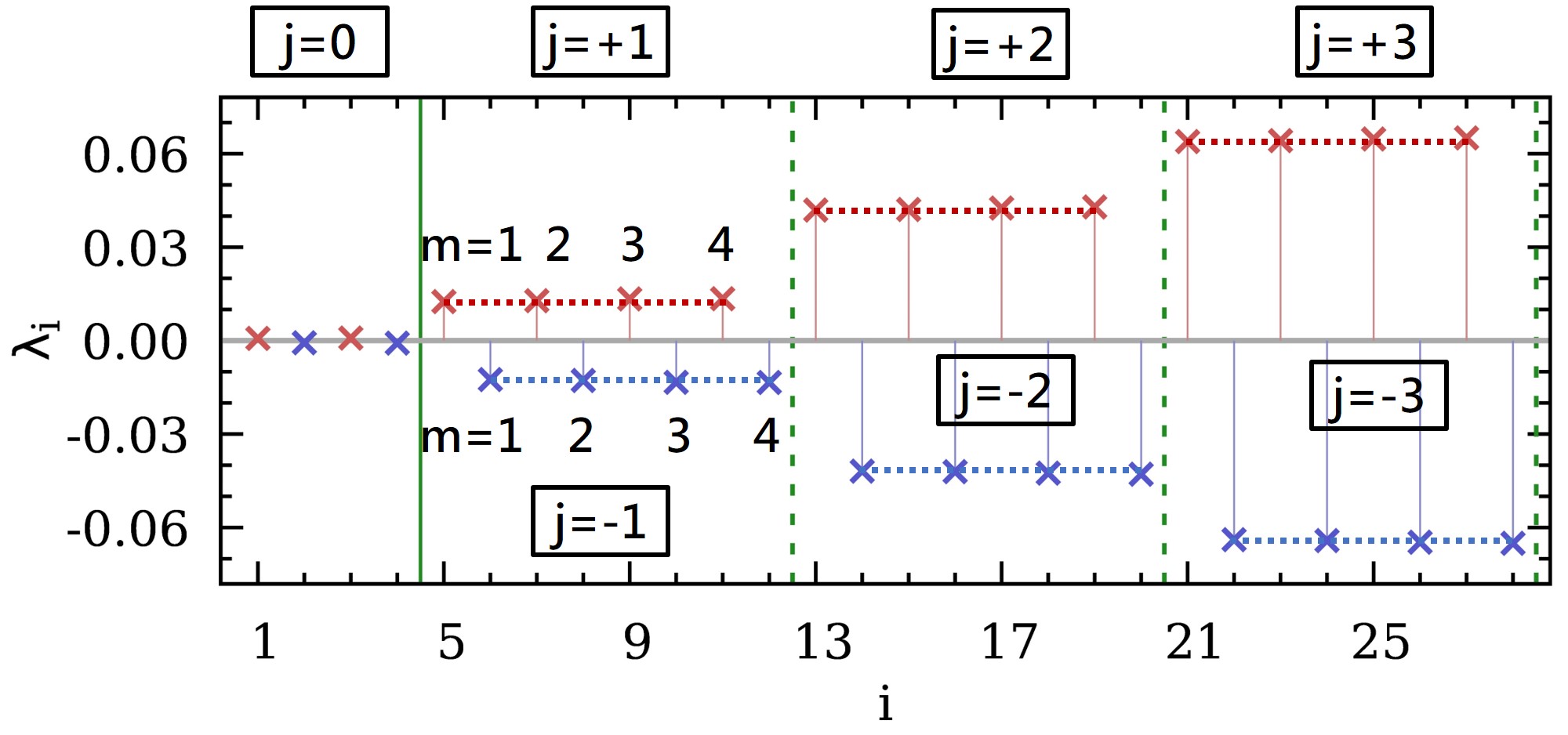}
  \caption{Staggered Dirac eigenvalue spectrum.}
  \label{fig:quartet}
\end{figure}
Let us describe the serial notation for Dirac eigenmodes.
So far we have used a serial index $i$ for an eigenvalue $\lambda_i$
which is assigned in ascending order such that $i < j$ if $\lambda_i <
\lambda_j$.
It is convenient since we can use a compact notation such as
$\Gamma_5^{i,k}$.
Since the $SU(4)$ taste symmetry breaking effect is so small at $a \ne
0$, we can also introduce the quartet notation which reflects on the
taste symmetry as in Figure~\ref{fig:quartet}.
Here, we adopt the same quartet notation as in Ref.~\cite{
  SWME:2020yip}:
\begin{equation}
  D_s\ket{f_{j,m}} = i\lambda_{j,m}\ket{f_{j,m}}
  \equiv i\lambda_{j,m}\ket{j,m}
  \label{eq:eigen_eq_qurtet}
\end{equation}
where $j$ is a quartet index and $m$ is the index of a quartet
component ($m=1,2,3,4$).
In the continuum ($a = 0$), $\lambda_{j,m} = \lambda_{j,n}$ for $m \ne
n$ thanks to the $SU(4)$ taste symmetry.
But on the lattice ($a > 0$), $\lambda_{j,m} \ne \lambda_{j,n}$ for $m
\ne n$ as one can see in Fig.~\ref{fig:quartet}.
The ono-to-one mapping between a serial index $i$ and a quartet index
$(j,m)$ is given in Table~\ref{tab:1to1_map} for topological charge
$Q=\pm 1$.
Here, note that $\lambda_{2n-1} = -\lambda_{2n}$ for $n > 0$, and
$\lambda_{j,m} = -\lambda_{-j,m}$.
%
\begin{table}[h!]
  \centering
  \setlength{\tabcolsep}{2.5pt}
  \resizebox{1.00\linewidth}{!}{
    \begin{tabular}{|c|cccc|cccc|cccc|c|}
      \hline
      \hline
      $\lambda_i$       & $\lambda_{1}$ & $\lambda_{2}$ & $\lambda_{3}$
      & $\lambda_{4}$   & $\lambda_{5}$ & $\lambda_{7}$ & $\lambda_{9}$
      & $\lambda_{11}$  & $\lambda_{6}$ & $\lambda_{8}$ & $\lambda_{10}$
      & $\lambda_{12}$  & $\cdots$ \\ \hline
      $\lambda_{j,m}$   & $\lambda_{0,1}$   & $\lambda_{0,2}$ &
      $\lambda_{0,3}$   & $\lambda_{0,4}$   & $\lambda_{+1,1}$  &
      $\lambda_{+1,2}$  & $\lambda_{+1,3}$  & $\lambda_{+1,4}$  &
      $\lambda_{-1,1}$  & $\lambda_{-1,2}$  & $\lambda_{-1,3}$  &
      $\lambda_{-1,4}$  & $\cdots$ \\ \hline
      $i$               & $1$ & $2$ & $3$ & $4$ & $5$ & $7$ & $9$ & $11$
      & $6$ & $8$ & $10$  & $12$ & $\cdots$  \\ \hline
      $(j,m)$           & $(0,1)$ & $(0,2)$ & $(0,3)$ & $(0,4)$ &
      $(+1,1)$  & $(+1,2)$  & $(+1,3)$  & $(+1,4)$  & $(-1,1)$  &
      $(-1,2)$  & $(-1,3)$  & $(-1,4)$  & $\cdots$ \\ \hline
      type & \small Z  &  \small Z  &  \small Z  & \small Z  & \small NZ
      &  \small NZ & \small NZ & \small NZ & \small NZ & \small NZ &
      \small NZ & \small NZ & $\cdots$ \\
      \hline
      \hline
    \end{tabular}
  }
  \caption{One-to-one correspondence between serial index $i$ and
    quartet index $(j,m)$ for $Q=\pm 1$. Here, Z (NZ)
    represents zero (non-zero) modes.}
  \label{tab:1to1_map}
\end{table}

\begin{figure}[t!]
  \vspace*{-1mm}
  \begin{subfigure}{0.5\linewidth}
    \includegraphics[width=\linewidth]{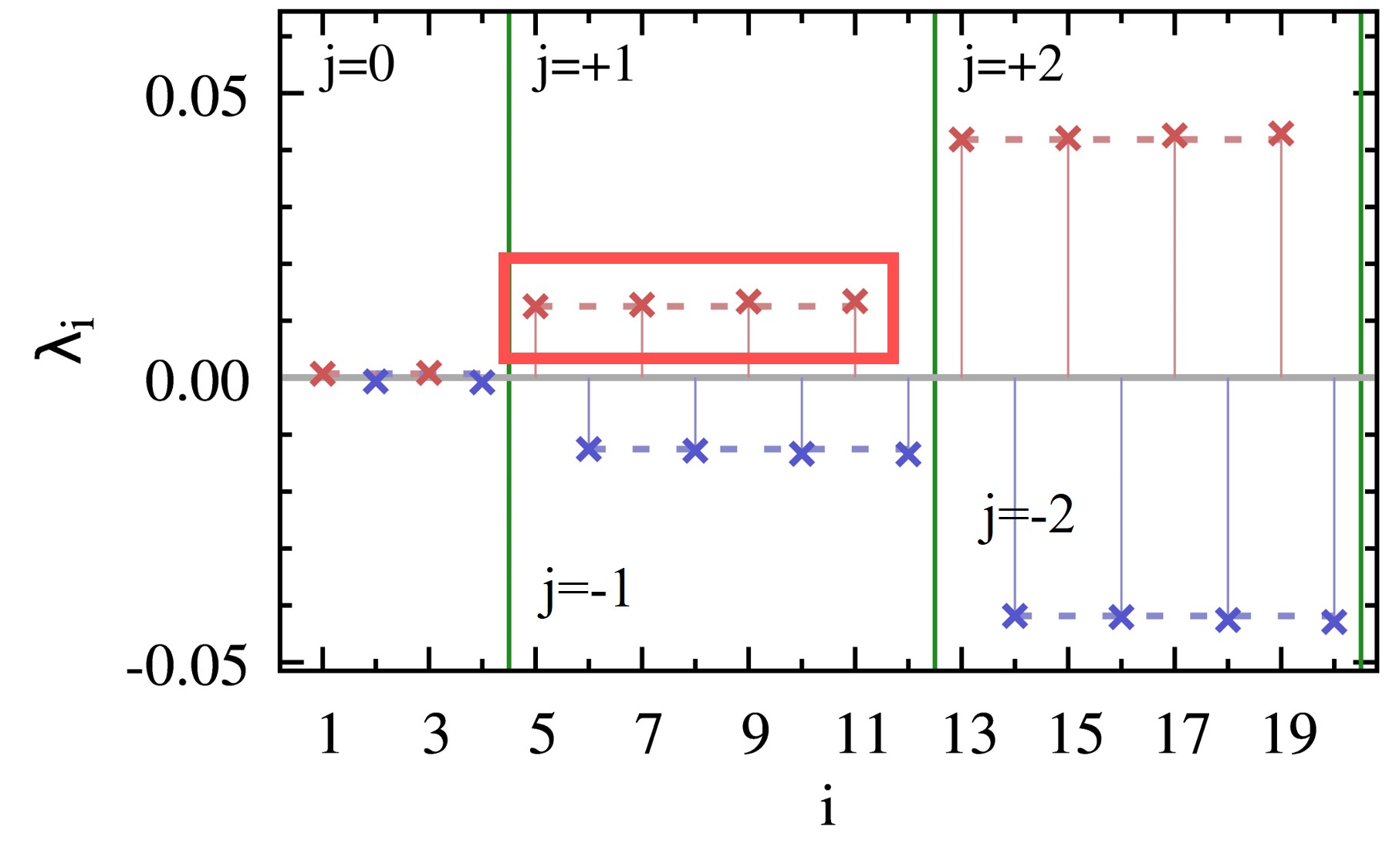}
    \caption{A quartet structure}
  \end{subfigure}
  \hfill
  \begin{subfigure}{0.5\linewidth}
    \includegraphics[width=\linewidth]{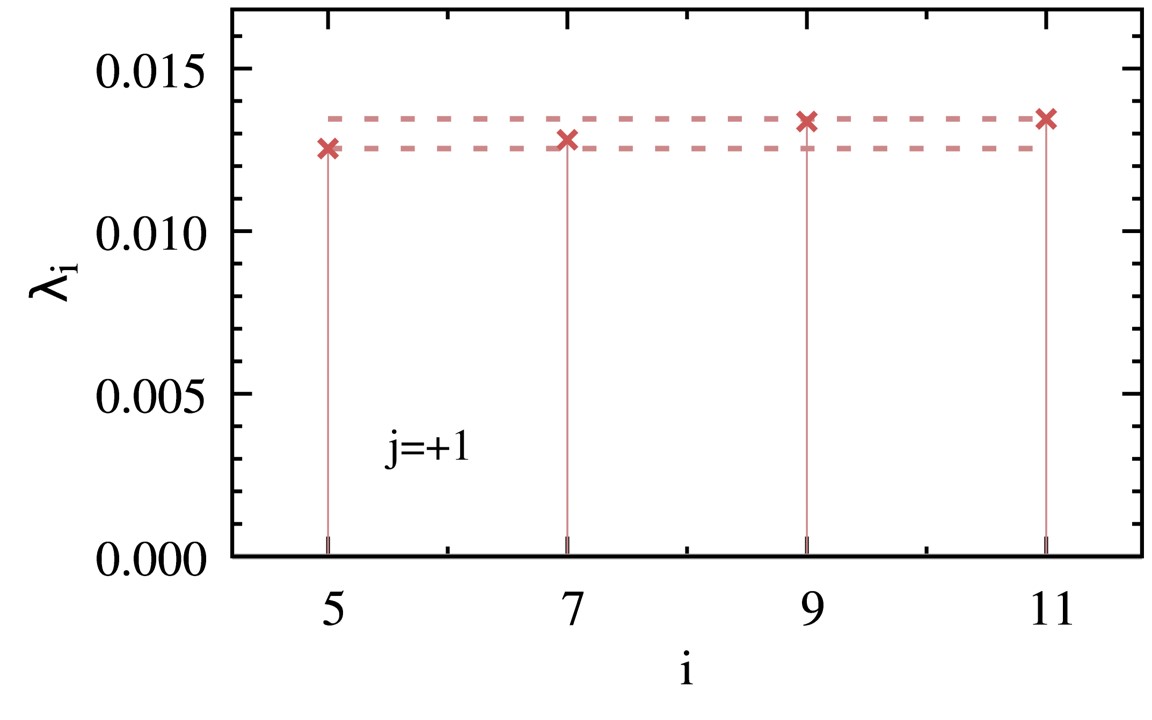}
    \caption{A quartet structure for the smallest non-zero mode.}
  \end{subfigure}
  \caption{Examples for the quartet structure.}
  \label{fig:approx_quartet}
\end{figure}

We find a minute deviation between eigenvalues within a quartet,
$\lambda_{j,m} \neq \lambda_{j, n}$ for $m \ne n$ since the taste
symmetry breaking is so tiny.
We find that the size of taste symmetry breaking is only 7\% even for
the smallest non-zero quartet as one can see in
Fig.~\ref{fig:approx_quartet}.
%
%
%

%
\begin{table}[!t]
  \centering
  \resizebox{0.85\linewidth}{!}{
  \begin{tabular}{r|l}
    \hline
    \hline
    parameters                      & values \\
    \hline
    gluonic action                  & Tree level Symanzik and tadpole
		improvement \cite{Luscher:1984xn, Alford:1995hw, Lepage:1992xa} \\
    lattice geometry                & $20^4$ \\
    $a$                             & $0.077(1)$ fm \\
    $\beta$     & $5.0$ \\
    $N_f$                           & 0 (quenched QCD) \\
    number of gauge configurations  & $300$ \\
    \hline
    valence quarks                  & HYP staggered fermions \cite{Hasenfratz:2001hp} \\
    \hline
    \hline
  \end{tabular}
	}
  \caption{Details on gauge configuration \cite{Follana:2005km}.}
  \label{tab:gauge_conf}
\end{table}
%
%

%
%
%

%
\section{Leakage pattern}
%
\begin{figure}[t!]
	\vspace*{-7mm}
  \centering
    \includegraphics[width=0.52\linewidth]{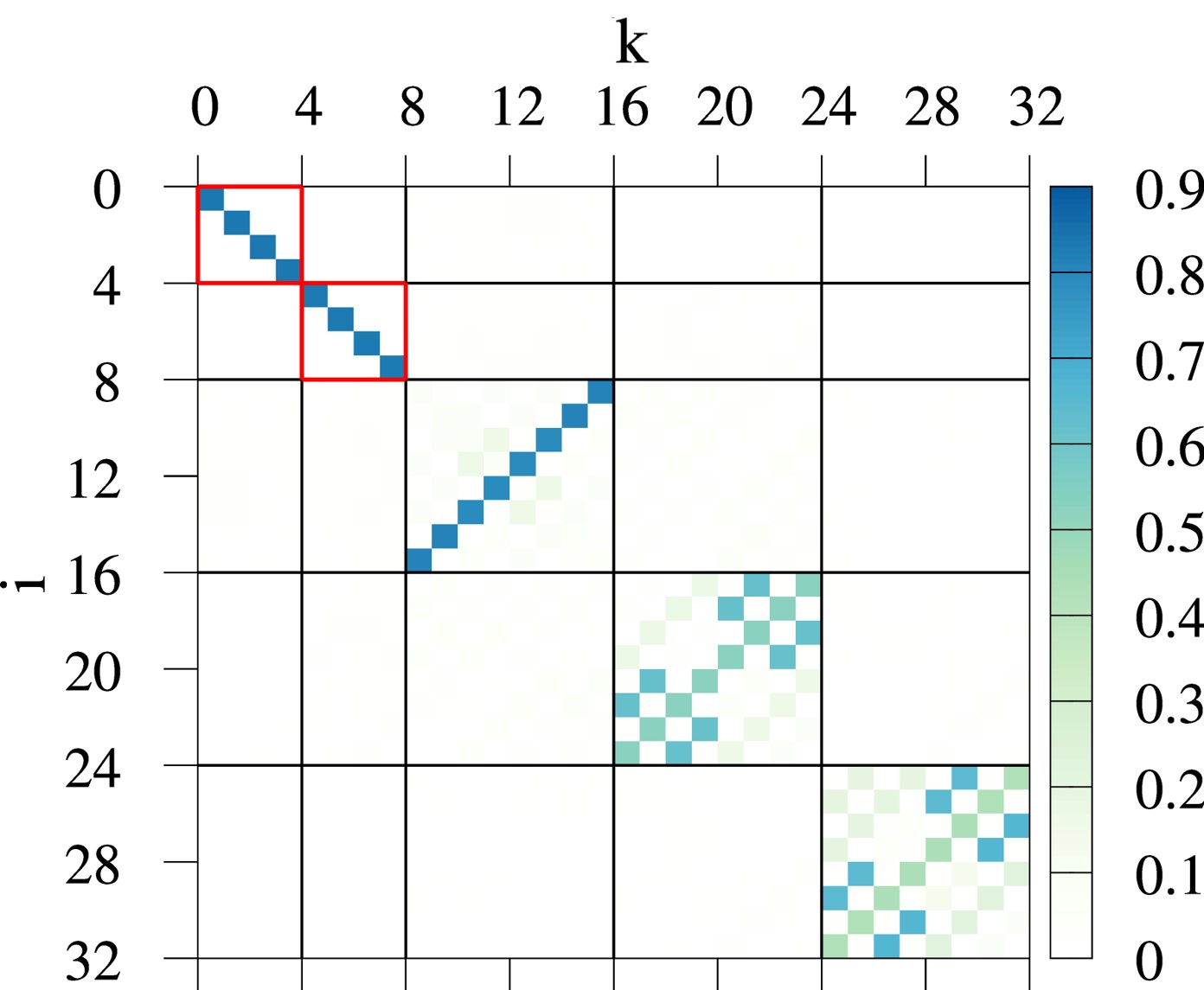}
	\caption{An example of leakage patterns.}
  \label{fig:leakage_pattern}
\end{figure}
First, let us consider the continuum case ($a=0$).
A shift operator, which is also a generator for the
$SU(4)$ taste symmetry, is
\begin{equation}
  \Xi_F = [\mathbb{1} \otimes \xi_F]
\end{equation}
where $\xi_F \in \{\xi_5, \xi_\mu, \xi_{\mu 5}, \xi_{\mu \nu}\}$
(for $\mu \neq \nu$).
Here, $\xi_\mu$ satisfies the Clifford algebra $\{\xi_\mu, \xi_\nu\} =
2\delta_{\mu \nu}$.
Since the taste symmetry is conserved, $\Xi_F$ commutes with staggered
Dirac operator,
\begin{equation}
  0 = \mate{f_i}{[D_s, \Xi_F]}{f_k} 
  = i(\lambda_i -	\lambda_k)|\Xi_F^{i,k}|~.
\end{equation}
If $\lambda_i \neq \lambda_k$, $|\Xi_F^{i,k}| = \mate{f_i}{\Xi_F}{f_k}
= 0$.
If $\lambda_i = \lambda_k$, $|\Xi_F^{i,k}| \neq 0$ is
possible.
$|\Xi_F^{i,k}|^2$ is a probability that one eigenstate transforms
into another by $\Xi_F$.
Hence, we call $|\Xi_F^{i,k}|$ the leakage parameter for the $\Xi_F$
operator.
Here, note that the leakage by $\Xi_F$ can occur only within a
quartet.
And among the set of $\{\Xi_F\}$, $\Xi_5$ is especially important
because it is related to $\Gamma_5$ by the Ward identity
in Eqs.~\eqref{eq:wi-2} and \eqref{eq:wi-3}.
Hence, $|\Gamma_5^{jm,-jn}|$ is just a mirror image of
$|\Xi_5^{jm,jn}|$ for the non-zero modes ($j \ne 0$).
Therefore, in the continuum ($a=0$),
\begin{equation}
  \Gamma_5^{jm,\ell n}
  = \mate{j,m}{\gamma_5\otimes\mathbb{1}}{\ell, n}
  = \begin{cases}
    ~\delta_{j\ell}\,\delta_{mn} 
    &\text{for zero modes,} \\
    ~\delta_{j,-\ell}\,\delta_{mn}
    &\text{for non-zero modes.}
  \end{cases} 
\end{equation}
On the lattice ($a \ne 0$),
\begin{equation}
  \Gamma_5^{jm,\ell n}
  =\mate{j,m}{\gamma_5\otimes\mathbb{1}}{\ell, n}
  \simeq 
  \begin{cases}
    ~\delta_{j\ell}\,\delta_{mn} 
    &\text{for zero modes,} \\
    ~\delta_{j,-\ell}\,c_{mn}
    &\text{for non-zero modes,} 
  \end{cases}
\end{equation}
where $c_{mn} \ne 0$ due to the taste symmetry breaking.
Hence, $|\Gamma_5^{i,k}|$ has a specific pattern which we call ``
leakage pattern''.
For zero modes it is diagonal within a quartet.
For non-zero modes it is off-diagonal and non-trivial only between a
quartet and its parity partner (we call them ``octet'' collectively
because the parity partner can be determined from a quartet
completely).
Therefore, zero modes and non-zero modes can be distinguished by the
leakage pattern.
For example, we show a typical leakage pattern in
Fig.~\ref{fig:leakage_pattern}.
Here, the gauge configuration has $Q=+2$; $n_-=2$ (left-handed zero
modes), $n_+=0$ (right handed zero modes).
The chroma of blue color represents a value of $|\Gamma_5^{i,k}|$.
Here, $i$ and $k$ are serial indices. 
The two red boxes represent two zero modes: one ($0 \le i \le 3$)
and the other ($4 \le i \le 7$).
We find that the leakage pattern for zero modes is diagonal within a
quartet.
There are three non-zero modes in Fig.~\ref{fig:leakage_pattern}.
One represents $j=\pm 1$ ($8 \le i \le 15$), another $j=\pm 2$ ($16
\le i \le 23$), and the other $j=\pm 3$ ($24 \le i \le 31$).
The leakage pattern of non-zero modes is off-diagonal and random
within an octet.

The leakage pattern (LP) method has a number of advantages compared with
the spectral flow (SF) method \cite{Adams:2009eb}. 
First, the LP method is as robust as the SF method in identifying
zero modes and non-zero modes.
Second, the LP method is, at least by a factor of thousand, cheaper
than the SF method in computational cost.
%
%

%
\section{Deep learning study}
Here, we want to address a question: \texttt{ Is the leakage pattern
  valid and universal over the entire gauge configurations? }
Since there are too many zero modes and non-zero modes with various
$Q$ values, it is practically impossible to check the leakage pattern
over the entire gauge configurations by analytic methods or by visual
examination.
Here, we introduce the deep learning (DL) techniques to solve the
problem, which recognize a specific pattern embedded in the data if it
exists, even though the data look completely random in the eye sight.
The DL method allows us to check the leakage pattern for non-zero
modes through the whole gauge configurations.

For the numerical study on the DL method, we use the lowest $200
\times 200$ matrix elements of $|\Gamma^{i,k}_5|$ ($0 \le i,k <200$)
on the lattice specified in Table~\ref{tab:gauge_conf}.

\subsection{Data sampling}
%
\begin{figure}[t!]
  \vspace*{-7mm}
  \centering
  \begin{subfigure}{0.45\linewidth}
    \includegraphics[width=\linewidth]{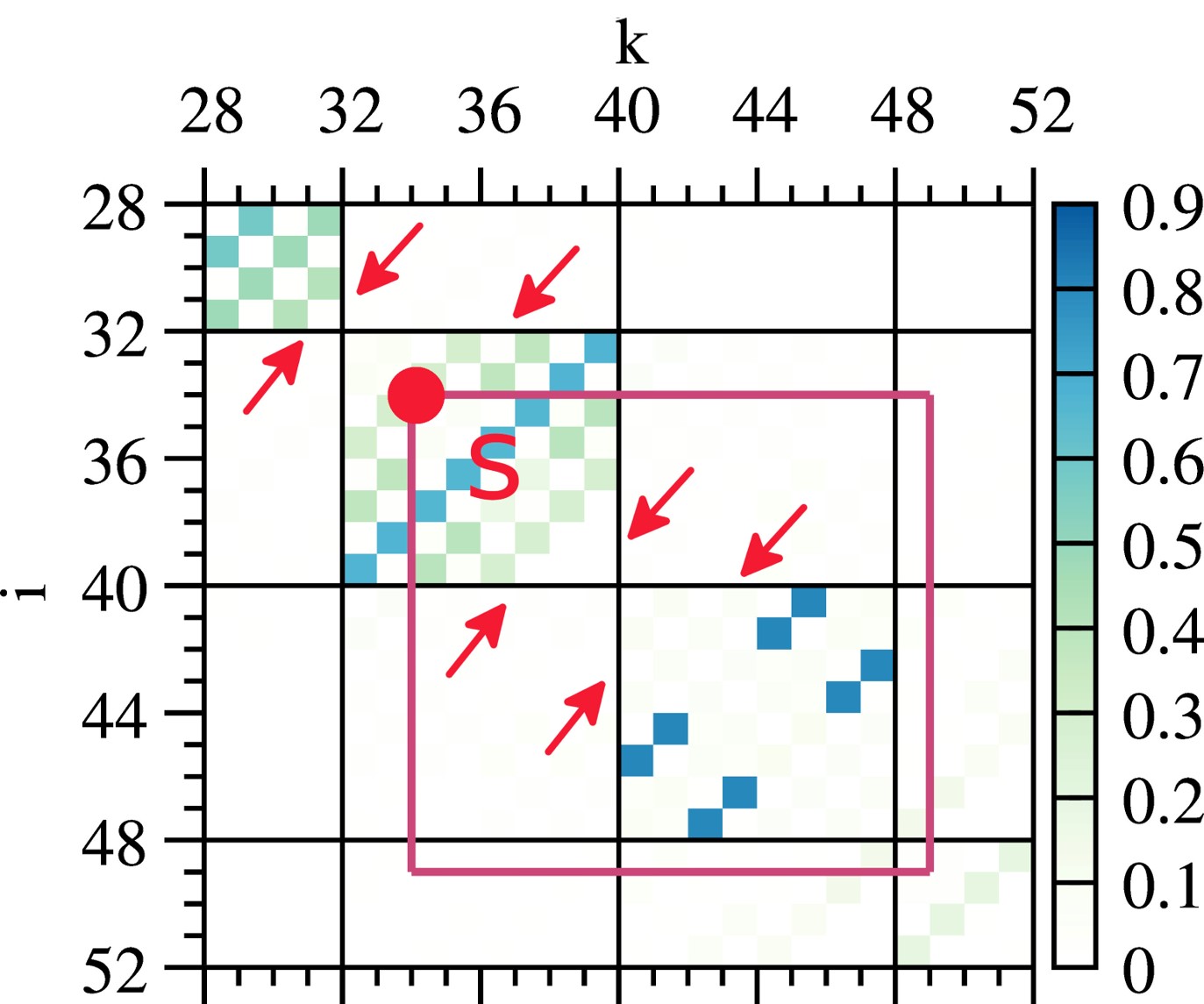}
    \caption{Example for data sampling.}
    \label{subfig:data_sampling}
  \end{subfigure}
  \hfill
  \begin{subfigure}{0.45\linewidth}
    \includegraphics[width=\linewidth]{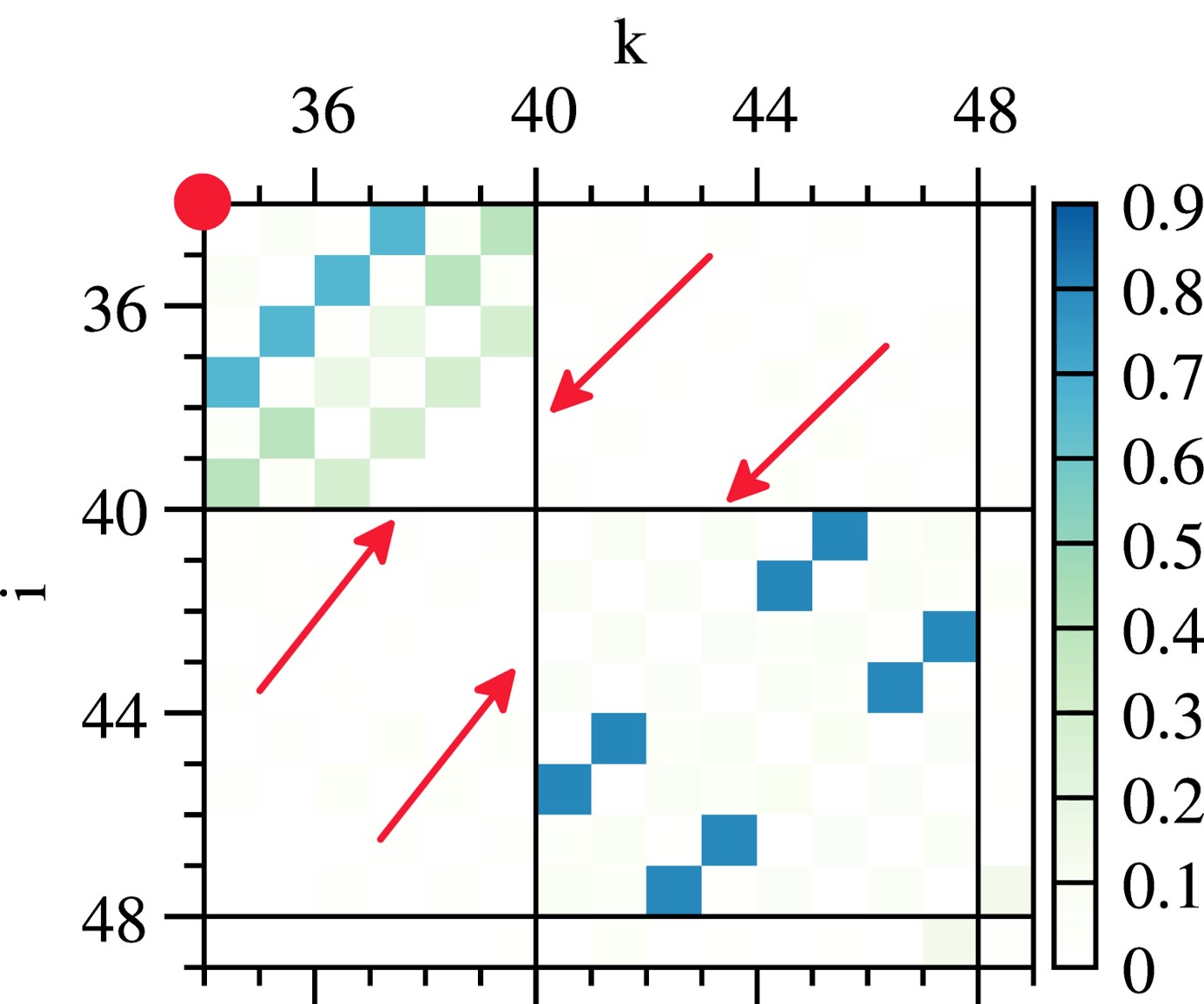}
    \caption{A $15\times 15$ data sample ($\red{s}=34$).}
    \label{subfig:data_sample}
  \end{subfigure}
  \caption{ (\subref{subfig:data_sample}) is a zoom-in version of the
    red square of (\subref{subfig:data_sampling}). The red square
    lines represent a typical data sample. The red blob in the upper
    left corner of the red square represents the sample serial index
    $s$ in Eq.~\eqref{eq:class-id}. The red arrows point to the black
    border lines between non-zero mode octets.}
  \label{fig:data_sampling}
\end{figure}
Here, our task belongs to a classification problem in DL.
We select $15\times 15$ sub-matrices randomly along the diagonal
($i=k$) line of the $200 \times 200$ matrix of $|\Gamma^{i,k}_5|$.
There is no overlap among the sub-matrices so that the probability
distribution should be independent and identical.
The sub-matrices are classified by the location of the border line
between two adjacent octets as in Fig.~\ref{fig:data_sampling}, where
an octet is composed of a quartet and its parity partner of non-zero
modes.
We choose the size of the sub-matrices as $15 \times 15$ from the
following guidelines.
\begin{itemize} \item We want to maximize statistics or the number of
samples, which requires minimizing the size of the sub-matrices.
\item We want to make the whole samples contain at least one complete
set of an octet.  \end{itemize}
The total number of different classes or labels (in terms of DL) is
eight, which is equal to the number of components in an octet.
We define the class ID, $C$ as follows,
\begin{equation}
  C = (s-i_b) \text{~mod } 8
  \label{eq:class-id}
\end{equation}
where $s$ is a serial index of the left-upper corner in a sub-matrix
(\textit{e.g.\ } $i=k=34$, the red blob in
Fig.~\ref{subfig:data_sample}), and $i_b$ is a serial index for the
border line (\textit{e.g.\ } $i=40$ and $k=40$, black lines in
Fig.~\ref{subfig:data_sample}).
Examples for the eight classes are presented in Fig.~$10$ of
Ref.~\cite{SWME:2020yip}.

\subsection{Deep learning analysis}
%
\begin{table}[!t]
 	\vspace*{-5mm}
  \centering
  \resizebox{0.90\linewidth}{!}{
  \begin{tabular}{ @{\qquad} l @{\qquad} | @{\qquad} l @{\qquad}}
    \hline\hline
    DL model structure & our specific choice \\
    \hline
    Loss function                            & categorical cross-entropy \\
    Hidden layer's activation function       & ReLU \\
    Output layer's activation function       & Softmax \\
    Optimization algorithm for loss function & Adam \\
    Neural network layer                     & MLP, CNN, CNN + MLP \\
    Auto-hyperparameter tuner                & Keras tuner (random search) \\
    \hline\hline
  \end{tabular}
	}
  \caption{DL model structure}
  \label{tab:DL_model_structure}
\end{table}
Details on the structure of our DL model are summarized in
Table~\ref{tab:DL_model_structure}.
Here, ReLU, MLP, and CNN represents rectified linear unit, multi layer
Perceptron, and convolutional neural network, respectively.
In order to perform the DL study, we need three types of data sets by
construction: one for training, another for validation, and the other
for test.
Details on the DL data sets are given in Table~\ref{tab:DL_data_set}.
%
\begin{table}[!t]
  \vspace*{-5mm}
  \begin{subtable}{0.40\linewidth}
    \centering
    \renewcommand{\arraystretch}{1.6}
    \resizebox{1.0\linewidth}{!}{
      \begin{tabular}{@{\quad} l @{\qquad} | @{\qquad} r @{\qquad} | @{\qquad} r @{\quad}}
	\hline \hline
	data set    & ngc & nds  \\
	\hline
	training    & 120 & 1223 \\
	validation  &  30 &  308 \\
	test        & 142 & 1441 \\
	\hline
	\hline
      \end{tabular}
    }
    \caption{DL data sets}
    \label{tab:DL_data_set}
  \end{subtable}
  \hfill
  \begin{subtable}{0.58\linewidth}   
    \centering
    \renewcommand{\arraystretch}{1.15}
    \resizebox{1.0\linewidth}{!}{
      \begin{tabular}{@{\qquad} l @{\qquad} | @{\qquad} r @{\qquad} r @{\qquad} r @{\qquad}}
	\hline
	\hline
	layer           & NN type & \# of units & act.f.\\
	\hline
	input           & -       & 225         & -         \\
	1st hidden      & MLP     & 160         & ReLU      \\
	2nd hidden      & MLP     & 1210        & ReLU      \\
	3rd hidden      & MLP     & 1490        & ReLU      \\
	output          & MLP     & 8           & softmax   \\
	\hline
	\hline
      \end{tabular}
    }
    \caption{The hyperparameters of the best performance model.}
    \label{tab:best_model}
  \end{subtable}
  \caption{ DL data sets and the best performance
    model. (\subref{tab:DL_data_set}) The ngc (nds) represents the
    number of gauge configurations (the number of data
    samples). (\subref{tab:best_model}) The NN denotes neural
    network and the act.f. activation function. }
\end{table}
When we perform the auto hyperparameter tuning, we find out that our
best performance model is MLP.
The hyperparameters for MLP are given in Table~\ref{tab:best_model}.
There are about ten samples ($15 \times 15$ sub-matrices) per gauge
configuration as shown in Table~\ref{tab:DL_data_set}.
We adopt the \texttt{accuracy} metric to measure the performance of DL
model.
The accuracy is defined as a ratio $TC/TND$, where $TC$ represents the
number of truly classified samples (DL prediction is equal to true
class), and $TND$ represents the total number of the data samples. 
In the test data set, we find that there are two independent sets of
gauge configurations which are distinct by the accuracy: one set of
five gauge configurations has accuracy less than 0.5, and the other
set of 137 gauge configurations has accuracy of about 0.9.
The former are called ``abnormal set'' and the latter
``normal set''.
The abnormal gauge configurations are entirely caused by the ghost
eigenvalue problem of Lanczos algorithm \cite{CULLUM198537}.
Before we begin our numerical study on the DL method, we have excluded
by hand those abnormal gauge configurations on which the topological
charge $Q$ measured by the index theorem is different from that of the
direct measurement.
In fact, the DL method is the best to identify the abnormal gauge
configurations such that we can filter them out.

Let us consider the accuracy $A_i$ measured on a gauge configuration
with an index $i$.
Then we can consider a set of the accuracies $S_a = \{A_i\}$ for
$1 \le i \le N$.
We also have a set of the number of samples $M_i$ for each gauge configuration
$S_n =\{ M_i\}$.
Then we can consider an weighted average of the accuracy $\bar{A}$
over the gauge configurations:
\begin{align}
  \bar{A} &= \frac{\sum\limits_i M_i A_i}{ \sum\limits_i M_i} \,,
  & &
  \sigma_{\bar{A}}^2 =
  \frac{\sum\limits_i M_i (A_i - \bar{A})^2}{ (N-1)\sum\limits_i M_i}
  \,,
\end{align}
where $\sigma_{\bar{A}}$ is the statistical error for $\bar{A}$. 
In Table \ref{tab:accuracy}, we present results for $\bar{A}$ measured
over the entire set (mixed = normal + abnormal), a subset of the
abnormal gauge configurations (abnormal), and a subset of the normal
gauge configurations (normal).
The accuracy for the abnormal subset is so small that the DL method
can identify it easily.
We find that the large error for the mixed set comes from that for the
abnormal subset.
We also find that the extremely high accuracy for the normal subset
reflects on the fact that the DL method succeeds in classifying the
samples of the normal subset, which leads to the key point that if we
select the majority of the DL prediction on each gauge configuration,
we can distinguish all the non-zero mode octets perfectly over the
entire normal subset.

We apply the DL method on the non-zero modes, because they are more
appropriate for the DL analysis due to the following reasons: 1)
abundant data, 2) complex structure, and 3) randomness in leakage
pattern.
The DL analysis gives a border line of the non-zero modes octets
robustly.
The leakage pattern for zero modes combined with the DL method is
as robust as the SF method in determining the topological charge $Q$.
However the key point is that the computational cost for the DL method
is much cheaper than the SF method by at least a factor of 1000.
%
%
\begin{table}[t!]
	\vspace*{-4.2mm}
  \centering
  \begin{tabular}{@{\qquad}l@{\quad}|@{\quad}r@{\quad}|@{\quad}l@{\qquad}}
    \hline\hline
    configuration & \# conf. & $\bar{A}$ [\%]\\
    \hline
    mixed          & 142     & $96.5(13)$ \\
    abnormal       & 5       & $17.3(78)$ \\
    normal         & 137     & $99.4(2)$  \\
    \hline\hline
  \end{tabular}
  \caption{Results for the accuracy $\bar{A}$ for mixed, abnormal, and
    normal set of gauge configurations in the test data set.}
  \label{tab:accuracy}
\end{table}

\subsection{ROC curve}
In the previous subsection, we use the accuracy metric for the DL analysis.
In general, for a skewed probability distribution of classes, the
accuracy is not such a good metric that it might give a misleading
information on the DL model performance \cite{10.1145/2907070}.
Since the class distribution of our data samples is not skewed but
uniform, the accuracy is a good metric for our DL analysis.
There are tens of alternative metrics in the DL market.
Among them, the AUC (\red{a}rea \red{u}nder ROC \red{c}urve)
\cite{FAWCETT2006861} is one of the most popular metrics, and can be
used instead of accuracy even if the class distribution is skewed.
Here, the ROC stands for \red{r}eceiver \red{o}perating
\red{c}haracteristic. 
In order to crosscheck results from the accuracy metric, we adopt
the AUC metric, even though we expect that both metrics will give
the same answer for our data sets in the end of day.
%

\begin{figure}[t!]
	\vspace*{-7mm}
  \centering
  \includegraphics[width=0.85\linewidth]{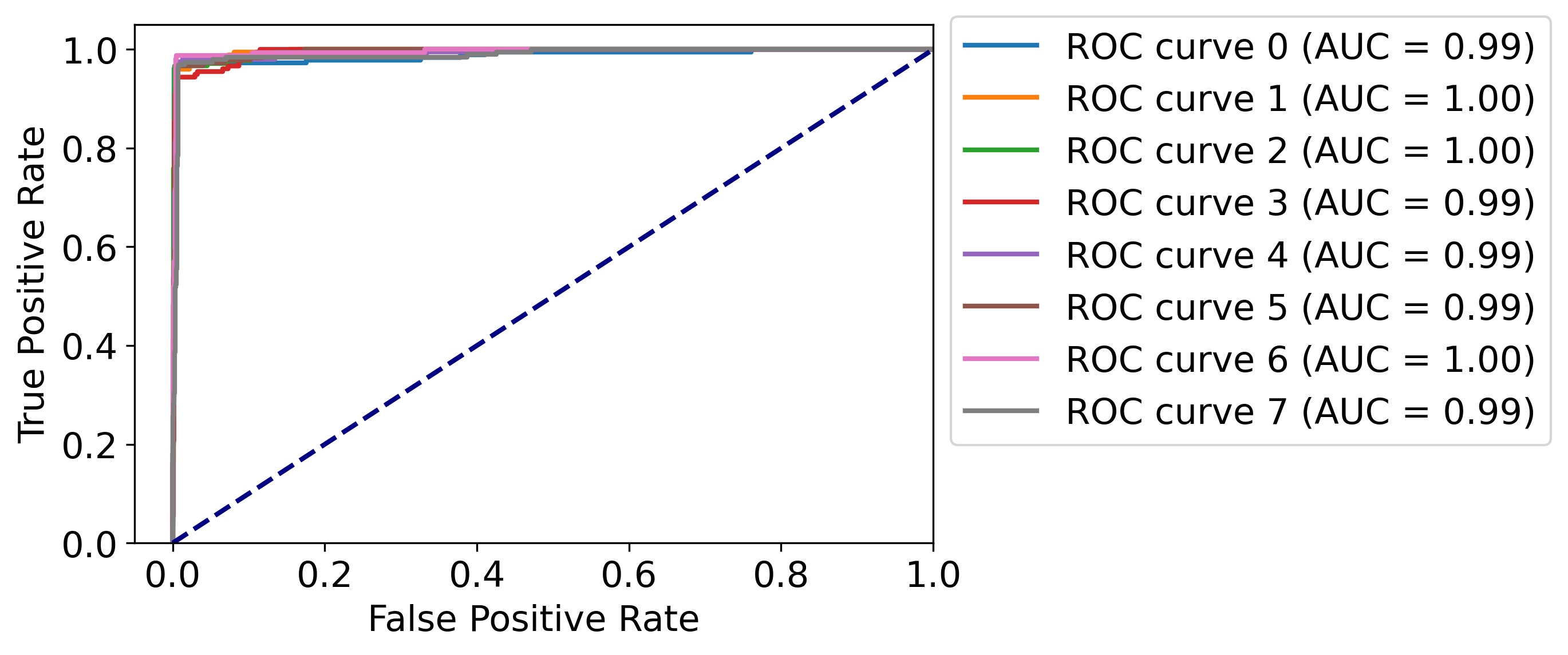}
  \caption{ROC curves and AUC metric for our best performance model}
  \label{fig:AUC}
\end{figure}

A ROC curve is defined in binary class (positive and negative classes).
As in Fig.~\ref{fig:AUC}, $y$-axis of ROC curve is true positive rate
(TPR) and $x$-axis is false positive rate (FPR).
\begin{align}
  \text{TPR} &= \frac{TP}{P} && \text{FPR} = \frac{FP}{N}
\end{align}
Here, $TP$ ($FP$) represents the number of the samples for which the
DL prediction is positive and their actual class is positive (negative).
The $P$ ($N$) is the total number of samples in the actual positive
(negative) class.
Here, the DL binary model gives a probability that each sample belongs to
the positive class.
The DL model prediction for each sample is determined by a threshold
applied to the probability to accept it (positive prediction) or
reject it (negative prediction).
The number of samples in the positive prediction changes according to
the threshold value, and so do the $TP$ and $FP$, which leads to
a corresponding change in TPR and FPR.
Hence, we may obtain the ROC curve for the DL binary model by running
the threshold from 0 to 1.
%

If we set the threshold to zero, all the predictions by the DL binary
model are positive, which results in a trivial case that $TP = P$ and
$FP = N$, which corresponds to $(1,1)$ in the ROC curve.
If the threshold is set to 1, all the preditions are negative, which
makes $TP=0$ and $FP=0$, which corresponds to $(0,0)$ in the ROC
curve.
Hence, the ROC curves, in general, share the two points $(1,1)$ and
$(0,0)$ in common.
If the DL predictions are perfect (\textit{i.e.} $TP=P$ and $FP=0$),
then the ROC curve passes through $(1,0)$.
AUC is the area under the ROC curve, and so AUC for a perfect DL model
is 1.
If the DL predictions are random (\textit{i.e.} $TP=0.5P$ and $FP=0.5N$),
the ROC curve corresponds to the blue dashed line in Fig.~\ref{fig:AUC},
which makes AUC=0.5.
Hence, AUC is an alternative good metric for the DL model performance.
If the DL model prediction gets better, its AUC becomes closer to 1. 
If the DL model prediction gets worse, its AUC becomes closer to 0.5. 
%

Since our DL model belongs to the multiclass classification, we need
to extend the binary-class picture of the ROC curve to the multiclass
picture.
We use the one-versus-rest method to extend the ROC curve to
the multiclass classification \cite{ FAWCETT2006861,
  Provost00well-trainedpets:}.
The one-versus-rest method converts multiclass classification into
binary class classification regarding $c_i$ as a positive class and
the rest ($\{c_j | j \ne i\}$) as a negative class.
In Fig.~\ref{fig:AUC}, we present the ROC curve and its AUC($c_i$) for
each class $c_i$ in our best DL model.
Our total AUC is 
\begin{equation}
  \text{AUC}_\text{tot} = \frac{\sum\limits_{c_i \in\, C}
	\text{AUC}(c_i) w(c_i)}{ \sum\limits_{c_i \in\, C} w(c_i)} 
\end{equation}
where $C$ is a set of all classes, and $w(c_i)$ is a weight of $c_i$
(= the number of samples in $c_i$).
Our best performance model (MLP) gives $\text{AUC}_\text{tot}=0.998
\simeq 1$ for the normal set.

The $\text{AUC}_\text{tot}$ value implies that our best DL model
works almost perfectly.
Results for the AUC metric are highly consistent with those for the
accuracy metric in the previous subsection.
Its physical meaning is that the leakage pattern for zero and non-zero
modes is universal over the normal gauge configurations.
%
%

%
\section{Conclusions}
%
%
Thanks to the $U(1)_A$ symmetry and the $SU(4)$ taste symmetry, the
leakage pattern (LP) for zero modes is quite different from that for
non-zero modes.
We use the deep learning (DL) method to verify that the LPs for the
chirality operator are universal over the normal gauge configurations,
which we can not prove analytically nor visually.
We find that, using the LP method combined with the DL method, we can
determine topological charge as robustly as the spectral flow (SF)
method.
Since the computational cost for the LP/DL method is much cheaper than
that for SF at least by a factor of 1000, the LP/DL method is highly
promising.

\acknowledgments

The research of W.Lee is supported by the Mid-Career Research Program Grant
[No.~NRF-2019R1A2C2085685] of the NRF grant funded by the Korean government
(MOE). This work was supported by Seoul National University Research Grant
[No.~0409-20190221]. W. Lee would like to acknowledge the support from the
KISTI supercomputing center through the strategic support program for the
supercomputing application research [KSC-2017-G2-0009, KSC-2017-G2-0014,
KSC-2018-G2-0004, KSC-2018-CHA-0010, KSC-2018-CHA-0043, KSC-2020-CHA-0001].
Computations were carried out in part on the DAVID supercomputer at Seoul
National University.

\bibliography{ref}

\end{document}